\documentclass{iopjournal}

\usepackage{ragged2e}
\usepackage{amsmath}
\justifying
\usepackage{cite}
\usepackage{amssymb}

\newcommand{\td}[0]{\mathrm{d}}

\begin{document}

\articletype{Paper} %	 e.g. Paper, Letter, Topical Review...

\title{Analytical estimates of core optics stray light noise and design requirements for ground based gravitational wave detectors}

\author{M. Andrés-Carcasona$^{1,2}$\orcid{0000-0002-8738-1672}, M. Evans$^1$\orcid{0000-0001-8459-4499}}

\affil{$^1$LIGO Laboratory, Massachusetts Institute of Technology, Cambridge, MA 02139, USA}

\affil{$^2$Kavli Institute for Astrophysics and Space Research, Massachusetts Institute of Technology, Cambridge, MA 02139, USA}

\email{mandresc@mit.edu}


\begin{abstract} 
\justifying 
 We present an analytical framework to translate detector sensitivity into stray light requirements for the core optics region of ground based gravitational wave interferometers. We derive analytical estimates for several relevant scattering regimes, extend the treatment to collections of planar baffles, and obtain an analogous requirement for deterministic ghost beams intercepted by beam dumps. The framework is intended to provide conservative design requirements before a complete optical layout is available. We apply it to Cosmic Explorer to show how these estimates can be used to derive requirements for next generation detectors. The resulting expressions provide a compact set of tools for stray light budgeting and early design studies in current and future gravitational wave detectors. 
\end{abstract}

\section{Introduction}
\label{sec:introduction}

Ground based gravitational wave (GW) detectors rely on the measurement of extremely small changes in the optical path length of kilometre scale laser interferometers. Their sensitivity depends on the level of many sources of noise, some of which will become increasingly important for third-generation (3G) observatories such as Cosmic Explorer (CE)~\cite{Evans:2021gyd,CE1,CE2,CE3} or Einstein Telescope (ET)~\cite{ETdesign,ETcds}, due to the expected improved strain sensitivity. This imposes more stringent requirements on the design of the interferometer.

Stray light, also known as scattered light, is produced whenever a fraction of the main optical field is scattered, diffracted, clipped, or redirected through a parasitic reflection. This light becomes problematic when it interacts with a non-optical surface and recouples to a field sensed by the interferometer~\cite{Accadia:2010zzb,LIGO:2020zwl,Longo:2020onu,Longo:2021avq,Longo:2023vac}. The phase accumulated along the parasitic path depends on the position of the scattering surface, such that motion of that surface is transferred to the recoupled optical field. Its interference with the nominal field can then appear as an effective displacement noise in the GW readout. When the relative motion becomes comparable to or larger than the laser wavelength, this coupling can become nonlinear and low-frequency motion can be upconverted into the observing band through phase wrapping~\cite{Ottaway:2012oce,Canuel:2013suy}.

The mitigation of this noise normally combines several complementary strategies. The generation of stray light can be reduced through an improvement in the optical quality of the mirror or its coating and by reducing the clipping of the beam. Light that nevertheless leaves the nominal path can be intercepted using baffles or beam dumps, while the motion of the illuminated surfaces can be reduced through seismic isolation or by mounting them so that they move approximately together with the recoupling optic. The scattering properties of these surfaces must also be controlled so that the mitigation surfaces do not themselves become important scatterers. 

Previous studies have developed analytical and numerical descriptions of these processes, together with techniques to mitigate them, and applied them to specific scattering paths in LIGO~\cite{Thorne89,FlanaganThorne95_Diff,Flanagan94,Flanagan95, LIGO:2020zwl,McGowan:2026lat,LIGO:2024kkz,Helmling-Cornell:2023wqe,Glanzer:2022avx}, Virgo~\cite{Vinet96,Vinet97,Brisson98,Andres-Carcasona:2022imx,Ballester:2021bua,Virgo:2022ysc}, and proposed 3G detectors~\cite{Andres-Carcasona:2023qom,Andres-Carcasona:2025xwq,CE_backscattering,VajenteCEbaffles,Macquet:2022simsVirgo,Andres-Carcasona:2026prv,Andres-Carcasona:2026wbq}. Most of these studies have focused on the main arms, while here we focus on the core optics area. The core optics region encompasses the central interferometer, including the power recycling cavity, beam splitter, internal telescope optics, input test masses (ITM), and signal extraction cavity. Typically, it also includes the main arms, but we deliberately exclude those as they are being studied separately. Its stray light paths can involve much shorter propagation distances, tilted optics, vacuum chambers, connecting tubes, and optical tables, which require specifically designed baffles and beam dumps to absorb stray light and ghost beams. The corresponding design is also likely to evolve substantially during the conceptual and preliminary design phases of a new observatory.

The goal of this work is to develop a set of analytical expressions to quickly evaluate the strain noise in the core optics area. These estimates can be useful for current detectors and can also be used to set stray light requirements for 3G detectors. To make the derivations applicable to different proposed observatories, we retain the main geometrical and optical parameters explicitly. This makes it possible to identify the dominant scalings, compare alternative mitigation strategies, and update the resulting requirements as the design matures.

The resulting expressions are intended as conservative, first order estimates and do not replace more advanced techniques such as ray tracing or fast Fourier transform (FFT) tools. Their purpose is instead to expose the physical dependencies that control each scattering path, establish preliminary quantitative requirements, guide optical and mechanical trade studies, and identify the areas for which more detailed modeling or dedicated measurements are necessary. We illustrate this procedure using representative parameters for CE and discuss the implications for the isolation and optical properties of surfaces surrounding its core optics.

\section{Strain noise from recoupled light}
\label{sec:strain_noise}

Stray light becomes a source of noise when light leaving the main beam interacts with a secondary surface and a fraction of it recouples to the interferometer. The magnitude of the resulting noise is determined mainly by two different aspects: the fraction of power that completes the scattering path and recouples with the main field, and the motion of the surface that modulates the phase of this returning light. In this section we first describe the optical recoupling, then connect it to the equivalent detector displacement noise, and finally discuss the role played by the motion and isolation of the backscattering surface and by phase wrapping.

\subsection{Scattered-light recoupling}
\label{sec:scattered_recoupling}

Following the formalism developed by Thorne and Flanagan~\cite{Thorne89,Flanagan94,Flanagan95,FlanaganThorne95_Diff}, the fraction of power that recouples with the main beam can be estimated as
\begin{equation}
\epsilon^2=\frac{\delta P}{P}=\left(\frac{\lambda}{r}\right)^2\left(\frac{\td P}{\td \Omega_{\rm ms}}\right)^2\frac{\td P}{\td\Omega_{\rm bs}}\delta\Omega_{\rm ms},
\label{eq:recoupling}
\end{equation}
where $\delta P$ is the amount of power that recouples into the main beam, $P$ is the power of the main beam, $\lambda$ is the laser wavelength, and $r$ is the distance to the backscattering surface. The quantity $\td P/\td\Omega_{\rm ms}$ is the probability per unit solid angle for the optic to scatter light towards the backscattering surface, while $\td P/\td\Omega_{\rm bs}$ describes the probability per unit solid angle for the backscatterer to send the light back towards the optic. The solid angle $\delta\Omega_{\rm ms}$ corresponds to the region of the backscattering surface illuminated by the scattered light.

The factor $(\td P/\td\Omega_{\rm ms})^2$ accounts for the scattering from the emitting optic towards the secondary surface and, by reciprocity, the corresponding coupling of the returning light back into the original optical mode, where it recombines with the main field. The factor $(\lambda/r)^2$ represents the diffraction limited efficiency with which light returning from a surface located at a distance $r$ can recouple to the main optical mode. It is therefore not sufficient for scattered light simply to propagate back towards the interferometer as only the fraction that overlaps with the relevant optical mode contributes to $\epsilon^2$. A simple estimate based on the coupling of scattered light into a Gaussian mode, given in App.~\ref{app:mode_recoupling}, recovers this scaling in both the near- and far-field limits.

The scattering probabilities can generally be described using the bidirectional reflectance distribution function (BRDF), related to the probability per unit solid angle through
\begin{equation}
{\rm BRDF}_i(\theta)=\frac{1}{\cos(\theta_\perp)}\frac{\td P}{\td\Omega_i},
\label{eq:brdf}
\end{equation}
where $\theta$ is measured with respect to the specular reflection direction and $\theta_\perp$ with respect to the surface normal. We assume that the incident angle $\theta_i$ and the specularly reflected angle $\theta_s$ are equal, and use $\theta$ to characterize the deviation of the scattered light from this specular direction. For small scattering angles the distinction between ${\rm BRDF}$ and $\td P/\td\Omega$ is small because $\cos(\theta_\perp)\simeq1$, whereas at wide angles this geometrical factor can become important. In the following, when the orientation of a secondary backscattering surface is not specified, we conservatively use its BRDF as an upper bound on the corresponding differential scattering probability,
\begin{equation}
\frac{\td P}{\td\Omega_{\rm bs}}
={\rm BRDF}_{\rm bs}\cos(\theta_{{\rm bs},\perp})
\leq {\rm BRDF}_{\rm bs}.
\label{eq:brdf_conservative_bound}
\end{equation}
Thus, cosine factors retained explicitly in the analytical expressions describe the known geometry of the emitting optic, while omission of the corresponding factor for an unspecified backscattering surface provides a conservative estimate.

For the core optics we model the BRDF of the optical surfaces using the power law
\begin{equation}
{\rm BRDF}(\theta)=\frac{\alpha}{\theta^n},
\label{eq:brdf_powerlaw}
\end{equation}
where $\alpha$ determines the overall scattering level and $n$ its angular dependence. This parametrization is simple yet representative of the angular dependence measured for good quality optical surfaces over the range relevant for stray light calculations~\cite{Romero-Rodriguez:2022mje}. We keep both $\alpha$ and $n$ explicit in the following derivations so that the expressions can be adapted to different optical surfaces. The case $n=2$ will be particularly relevant because it provides a useful reference model and leads to simple analytical expressions for several of the geometries considered in Sec.~\ref{sec:analytical_scattering}, and is the scaling originally considered by Thorne~\cite{Thorne89}. The use of Eq.~\eqref{eq:brdf_powerlaw} for tilted core optics and its connection with the surface spatial power spectrum are discussed in App.~\ref{app:tilted_brdf}. In particular, for the reference $n=2$ case the leading small-angle term is independent of the incidence angle, while tilt enters only through subleading corrections. The same appendix also provides a total integrated scattering estimate as a consistency check on the assumed value of $\alpha$.

Equation~\eqref{eq:recoupling} can therefore be interpreted as the link between the optical properties of the two surfaces and the amount of stray light that ultimately affects the interferometer. The remaining question is how large $\epsilon^2$ can be before the phase modulation introduced by the moving backscatterer becomes relevant for the detector sensitivity.

\subsection{Detector response and maximum allowed coupling}
\label{sec:maximum_coupling}

The amount of recoupled power that can be tolerated depends on the displacement noise produced by the moving backscattering surface. We denote by $x_{\rm bs}$ the displacement amplitude spectral density (ASD) of the backscatterer and by $x_{\rm DARM}$ the equivalent displacement noise in the differential arm degree of freedom. Ref.~\cite{KevinSECL_DARM} derives the equivalent DARM displacement produced by the phase modulation of a parasitic optical field in the signal extraction cavity. The same relation applies to recoupled stray light once the scattered field has coupled back into the relevant interferometer mode. Since $\epsilon^2=\delta P/P$, the amplitude of the recoupled field is $\epsilon$ times that of the nominal field. A displacement $x_{\rm bs}$ of the backscattering surface produces a phase modulation $\delta\phi=2kx_{\rm bs}$, with $k=2\pi/\lambda$, and therefore the leading order field perturbation is proportional to $2k x_{\rm bs}\epsilon$. This is formally identical to the motion sidebands considered in Ref.~\cite{KevinSECL_DARM}, so the relation is
\begin{equation}
\left|\frac{x_{\rm DARM}}{x_{\rm bs}}\right|=\sqrt{\epsilon^2\frac{\pi}{\mathcal{F}_a}}\sqrt{1+\left(\frac{f}{f_a}\right)^2},
\label{eq:darm_bs}
\end{equation}
where
\begin{equation}
f_a=\frac{c}{4\mathcal{F}_aL_a}
\label{eq:arm_pole}
\end{equation}
is the arm cavity pole, $L_a$ is the arm length, and
\begin{equation}
\mathcal{F}_a=\frac{\pi}{1-r_i}
\label{eq:arm_finesse}
\end{equation}
is the finesse of the arm cavity. Here, $r_i$ denotes the ITM amplitude reflectivity.

Equation~\eqref{eq:darm_bs} provides the connection between the geometrical and optical recoupling calculated from Eq.~\eqref{eq:recoupling} and the resulting detector noise. In particular, it can be directly inverted to obtain the maximum recoupled power fraction compatible with a given displacement-noise target,
\begin{equation}
\epsilon_{\rm max}^2(f)=\frac{\mathcal{F}_a}{\pi}\left|\frac{x_{\rm DARM}(f)}{x_{\rm bs}(f)}\right|^2\left[1+\left(\frac{f}{f_a}\right)^2\right]^{-1}.
\label{eq:epsilon_max}
\end{equation}
This expression is useful because it separates the two parts of the problem. The left-hand side of the comparison, $\epsilon^2$, is determined by the optical scattering path and will be calculated from the BRDFs and geometry in Sec.~\ref{sec:analytical_scattering}. The maximum allowed value $\epsilon_{\rm max}^2$ is instead determined by the detector sensitivity and by the motion of the surface receiving the stray light.

In the following we use the detector DARM sensitivity itself as the upper level noise target, so that $\epsilon_{\rm max}^2$ represents the largest recoupling compatible with scattered-light noise remaining below the total detector sensitivity. If a future detector noise budget allocates only a fraction $\eta_{\rm SL}$ of the DARM displacement target to scattered light, the same expressions apply after the replacement $x_{\rm DARM}\rightarrow\eta_{\rm SL}x_{\rm DARM}$, corresponding simply to $\epsilon_{\rm max}^2\rightarrow\eta_{\rm SL}^2\epsilon_{\rm max}^2$.

The requirement is intrinsically frequency dependent because both the detector displacement sensitivity and the motion of the backscatterer depend on frequency. A scattering path must therefore satisfy
\begin{equation}
\epsilon^2\leq\epsilon_{\rm max}^2(f)
\label{eq:epsilon_frequency_requirement}
\end{equation}
throughout the frequency interval over which scattered-light noise is required to remain below the detector noise. Since the optical coupling $\epsilon^2$ associated with a fixed geometry does not itself carry the frequency dependence appearing in Eq.~\eqref{eq:epsilon_max}, the practical requirement is set by the most restrictive value of $\epsilon_{\rm max}^2(f)$ across that interval. This provides a single upper-level recoupling requirement that can subsequently be translated into limits on the BRDF, area, location, or orientation of the surfaces involved in a particular scattering path.

The dependence on $x_{\rm bs}$ in Eq.~\eqref{eq:epsilon_max} also shows that the optical and mechanical requirements cannot be considered independently. The same value of $\epsilon^2$ can be acceptable for a well-isolated surface but excessive for a surface that follows the ground motion. Conversely, improving the mechanical isolation of a potential backscatterer directly relaxes the required suppression of the optical coupling.

\subsection{Surface motion, isolation and phase wrapping}
\label{sec:motion_wrapping}

The displacement spectrum $x_{\rm bs}$ depends strongly on how the backscattering surface is mechanically attached to the detector. Surfaces rigidly attached to the vacuum system can approximately follow the seismic motion, whereas surfaces mounted on isolated optical tables or suspended structures can experience substantially smaller motion. For the analytical estimates considered here, a useful simple model for the seismic displacement ASD is
\begin{equation}
x_{\rm bs}(f)=\frac{A}{f^2},
\label{eq:seismic_model}
\end{equation}
where $A$ sets the overall level of motion. The value of $A$ is detector- and site-dependent and will be specified for the Cosmic Explorer application in Sec.~\ref{sec:CE} rather than fixed at this stage.

Mechanical isolation enters the requirement in a particularly simple way. If an isolation system reduces the displacement ASD of the backscattering surface by an amplitude factor $H_{\rm iso}$, such that
\begin{equation}
x_{\rm bs}^{\rm iso}(f)=\frac{x_{\rm bs}(f)}{H_{\rm iso}(f)},
\label{eq:isolation}
\end{equation}
then Eq.~\eqref{eq:epsilon_max} immediately gives
\begin{equation}
\left(\epsilon_{\rm max}^{\rm iso}\right)^2=H_{\rm iso}^2\epsilon_{\rm max}^2.
\label{eq:isolation_epsilon}
\end{equation}
Thus, an improvement in the displacement amplitude by a factor $H_{\rm iso}$ relaxes the allowed power coupling by $H_{\rm iso}^2$. In practice, the displacement entering Eq.~\eqref{eq:darm_bs} should be understood as the motion that changes the optical path of the scattered field. Surfaces that are mechanically connected to, or move coherently with, the relevant optic can therefore have a substantially smaller effective motion than an equivalent surface attached directly to the vacuum system.

The previous discussion implicitly assumes that the phase modulation induced by the backscatterer remains in the approximately linear regime. We denote the corresponding time domain displacement by $x_{\rm bs}(t)$, to distinguish it from the displacement ASD $x_{\rm bs}(f)$ introduced above. A displacement $x_{\rm bs}(t)$ changes the round trip propagation distance of a backscattered field by approximately $2x_{\rm bs}(t)$ and therefore changes its optical phase by
\begin{equation}
\Delta\phi_{\rm bs}(t)=\frac{4\pi}{\lambda}x_{\rm bs}(t).
\label{eq:phase_motion}
\end{equation}
For displacements much smaller than the wavelength, the corresponding modulation can be linearized and the scattered light noise approximately follows the displacement spectrum of the surface. However, when the motion becomes sufficiently large for the phase to evolve through a significant fraction of $2\pi$, this approximation breaks down. The optical phase repeatedly wraps and the coupling becomes nonlinear~\cite{Ottaway:2012oce}.

This phase-wrapping regime is particularly relevant for surfaces with large low-frequency seismic motion. Although most of their displacement may occur below the gravitational-wave observing band, the nonlinear dependence of the scattered field on $\Delta\phi_{\rm bs}$ transfers part of this motion to higher frequencies. Consequently, low-frequency motion can produce scattered-light noise well above the original mechanical frequencies, a process generally referred to as upconversion or phase-wrapping~\cite{Ottaway:2012oce,Canuel:2013suy}.

When phase wrapping is important, the unwrapped displacement ASD of Eq.~\eqref{eq:seismic_model} alone is therefore not sufficient to determine the in-band noise. To account for it, a time domain realization $x_{\rm bs}(t)$ of the displacement spectrum is generated and converted into the optical phase through Eq.~\eqref{eq:phase_motion}. For an isolated surface, this spectrum is the residual motion obtained after applying the isolation transfer of Eq.~\eqref{eq:isolation}, so that phase wrapping is always evaluated after the mechanical attenuation has been applied. The nonlinear scattered light response is represented by the effective displacement~\cite{Ottaway:2012oce}
\begin{equation}
x_{\rm bs}^{\rm eff}(t)=\frac{\lambda}{4\pi}\sin\left[\frac{4\pi}{\lambda}x_{\rm bs}(t)\right],
\label{eq:effective_wrapped_motion}
\end{equation}
whose ASD is subsequently calculated and used as the effective backscatterer motion in Eq.~\eqref{eq:epsilon_max}. In the small motion limit $x_{\rm bs}^{\rm eff}(t)\simeq x_{\rm bs}(t)$, while for larger motion the nonlinear transformation transfers a low frequency displacement into higher frequencies and can therefore tighten the maximum allowed optical coupling.

Isolation is consequently useful in two different ways. It directly reduces $x_{\rm bs}$ and therefore relaxes the optical requirement according to Eq.~\eqref{eq:isolation_epsilon}, but it can also substantially reduce phase wrapping, thereby increasing the margin. The quantitative importance of both effects depends on the seismic environment and on the mechanical configuration of the potential backscattering surface. In Sec.~\ref{sec:CE} we evaluate these effects for representative Cosmic Explorer parameters and use them to determine the maximum recoupling that can be tolerated for the different classes of surfaces considered in the following sections.

\section{Analytical estimates for core optics scattering}
\label{sec:analytical_scattering}

\begin{figure}[htbp]
    \centering
    \includegraphics[width=1.0\linewidth]{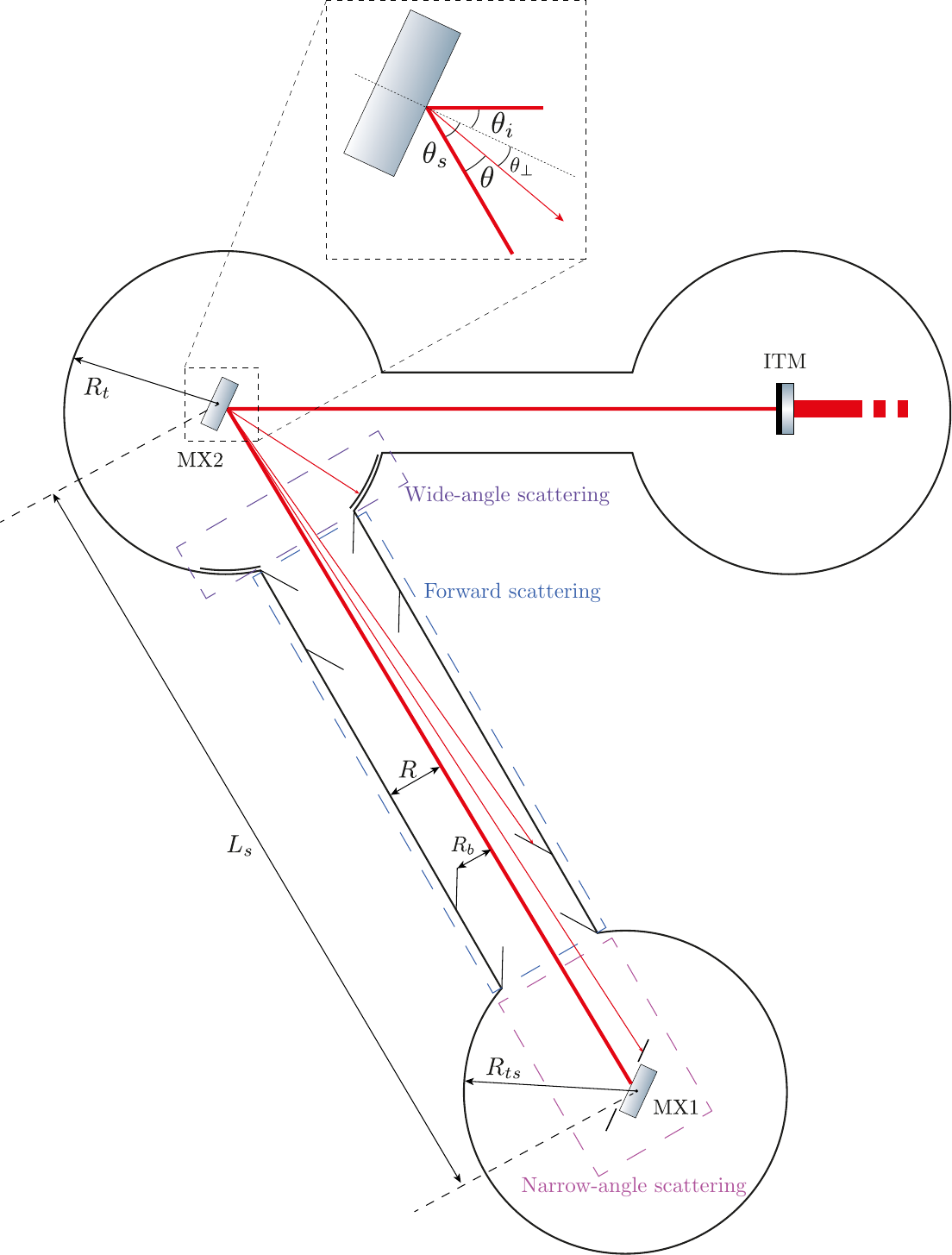}
    \caption{\justifying Representative core optics scattering geometry illustrating the narrow-angle, forward, and wide-angle regions considered in this work. The inset defines the scattering angle $\theta$ relative to the specular direction and the angle $\theta_\perp$ relative to the surface normal.}
    \label{fig:scattering_regions}
\end{figure}

We now use the recoupling formalism introduced in Sec.~\ref{sec:scattered_recoupling} to obtain analytical estimates for the main classes of scattering paths that can occur between core optics. Consider an emitting optic from which a fraction of the main beam is scattered towards a second, target optic. The surfaces that can intercept this light can be separated into three regions according to their position and scattering angle: narrow-angle scattering from surfaces close to the target optic, forward scattering from the tube or enclosure connecting the two optics, and wide-angle scattering from surfaces surrounding the emitting optic or otherwise located at large angles from the nominal propagation direction. This classification is geometrical rather than specific to a particular interferometer and can therefore be applied to different pairs of core optics. A representative geometry illustrating these three regions is shown in Fig.~\ref{fig:scattering_regions}. Although the example shown corresponds to a particular sequence of core optics, the same classification can be applied to any pair of optics connected by the main beam.

The same basic process occurs in all three cases. Light is first scattered by the emitting optic, interacts with a secondary surface, and a fraction is scattered back and recouples with the main optical mode. The differences between the regimes arise from the relation between the distance $r$, the scattering angle $\theta$, the solid angle subtended by the secondary surface, and the corresponding BRDF. We use the power-law model of Eq.~\eqref{eq:brdf_powerlaw} for the emitting optic and retain the BRDF of the secondary surface, ${\rm BRDF}_{\rm bs}$, explicitly. The resulting recoupling fractions can then be directly compared with the maximum value $\epsilon_{\rm max}^2$ derived in Sec.~\ref{sec:maximum_coupling}.

\subsection{Narrow-angle scattering}
\label{sec:narrow_scattering}

We first consider surfaces located close to the target optic. In this regime the scattering angle is small and the light reaching the surrounding surfaces can be conservatively approximated by the scattered intensity close to the edge of the target optic. For a backscattering surface of area $A_b$ located a distance $r$ from the emitting optic, the solid angle subtended by the surface is approximately
\begin{equation}
\delta\Omega_{\rm ms}\simeq\frac{A_b}{r^2}.
\label{eq:narrow_solid_angle}
\end{equation}
At small angles we can take $\td P/\td\Omega_{\rm bs}\simeq{\rm BRDF}_{\rm bs}$ and $\td P/\td\Omega_{\rm ms}\simeq\alpha/\theta^n$. Substitution into Eq.~\eqref{eq:recoupling} gives
\begin{equation}
\epsilon^2=\frac{\lambda^2}{r^2}{\rm BRDF}_{\rm bs}\frac{\alpha^2}{\theta^{2n}}\frac{A_b}{r^2}.
\label{eq:narrow_general}
\end{equation}

If $R_m$ denotes the radius of the target optic, the characteristic angle corresponding to light scattered towards its edge is $\theta\simeq R_m/r$. We conservatively use this value for the surfaces immediately surrounding the optic, rather than introducing an additional decrease of the scattered intensity with increasing distance from its edge. Equation~\eqref{eq:narrow_general} then becomes
\begin{equation}
\epsilon^2=\lambda^2\alpha^2{\rm BRDF}_{\rm bs}A_b\frac{r^{2n-4}}{R_m^{2n}}.
\label{eq:narrow_n}
\end{equation}
An interesting feature appears for the reference $n=2$ scattering model. In this case the explicit dependence on the distance cancels and the recoupled power becomes
\begin{equation}
\epsilon^2=\frac{\lambda^2\alpha^2}{R_m^4}{\rm BRDF}_{\rm bs}A_b.
\label{eq:narrow_n2}
\end{equation}
The decrease of the solid angle and mode recoupling efficiency with distance is exactly compensated by the increase in the main optic BRDF as the scattering angle approaches the specular direction. Within the approximations used here, moving a narrow-angle baffle farther from the emitting optic therefore does not reduce its coupling for a $\theta^{-2}$ BRDF. Instead, the relevant quantities are its illuminated area, its own BRDF, and the radius of the target optic.

Comparing Eq.~\eqref{eq:narrow_n2} with the maximum allowed coupling gives the general narrow-angle requirement
\begin{equation}
{\rm BRDF}_{\rm bs}A_b\leq\frac{R_m^4}{\lambda^2\alpha^2}\epsilon_{\rm max}^2.
\label{eq:narrow_requirement}
\end{equation}
This expression applies to baffles, chamber surfaces, or other elements surrounding a target optic as long as the small-angle approximation remains valid. Its numerical implication depends strongly on the motion of the particular surface through $\epsilon_{\rm max}^2$ and will be evaluated for representative Cosmic Explorer configurations in Sec.~\ref{sec:CE}.

\subsection{Forward scattering}
\label{sec:forward_scattering}

We next consider surfaces distributed along the region connecting the emitting and target optics. This includes, in particular, the walls of a vacuum tube and approximately axisymmetric annular baffling along it. Unlike the narrow-angle case, the scattering angle changes significantly along this region and the contribution has to be integrated over the corresponding angular interval. Discrete planar baffles that do not provide approximately azimuthally complete coverage are treated separately in Sec.~\ref{sec:planar_baffles}.

For an approximately axisymmetric tube wall, or for annular baffling treated in the continuum limit, an element subtends the solid angle
\begin{equation}
\td\Omega_{\rm ms}=2\pi\sin\theta\,\td\theta\simeq2\pi\theta\,\td\theta,
\label{eq:forward_solid_angle}
\end{equation}
where the final expression applies in the paraxial approximation. Assuming that the BRDF of the backscattering surface does not vary strongly over the relevant angular interval, Eq.~\eqref{eq:recoupling} gives
\begin{equation}
\epsilon^2=2\pi\lambda^2\alpha^2{\rm BRDF}_{\rm bs}\int_{\theta_{\rm min}}^{\theta_{\rm max}}\frac{1}{r^2}\theta^{1-2n}\td\theta.
\label{eq:forward_integral_r}
\end{equation}
For a tube with inner radius $R$, the distance from the emitting optic to the illuminated wall can be related geometrically to the scattering angle by $r\simeq R/\theta$. Substituting this relation gives
\begin{equation}
\epsilon^2=2\pi\left(\frac{\lambda\alpha}{R}\right)^2{\rm BRDF}_{\rm bs}\int_{\theta_{\rm min}}^{\theta_{\rm max}}\theta^{3-2n}\td\theta.
\label{eq:forward_integral}
\end{equation}
The recoupled power fraction is therefore
\begin{equation}
\epsilon^2=2\pi\left(\frac{\lambda\alpha}{R}\right)^2{\rm BRDF}_{\rm bs}
\begin{cases}
\dfrac{\theta_{\rm max}^{4-2n}-\theta_{\rm min}^{4-2n}}{4-2n}, & n\neq2,\\[8pt]
\ln\left(\dfrac{\theta_{\rm max}}{\theta_{\rm min}}\right), & n=2.
\end{cases}
\label{eq:forward_solution}
\end{equation}

The integration limits are fixed by the geometry of the two optical chambers and the connecting tube. For example, if $L_s$ is the separation between the optics, $R_t$ and $R_{ts}$ characterize the dimensions of the emitting and target chambers, respectively, and $R_b$ is the clear radius of the baffles along the tube, a representative geometry gives
\begin{equation}
\theta_{\rm min}\simeq\frac{R_b}{L_s-R_{ts}},\qquad \theta_{\rm max}\simeq\arctan\left(\frac{R}{R_t}\right).
\label{eq:forward_angles}
\end{equation}
The precise limits can be replaced by those corresponding to the actual detector geometry without modifying the derivation.

For $n=2$, the dependence on the angular limits is only logarithmic. Equivalently, equal logarithmic intervals in scattering angle contribute equally to the integral. The forward-scattering estimate is therefore comparatively insensitive to modest changes in the exact transition between the narrow-angle, forward, and wide-angle regions. The corresponding requirement on the backscattering surface is
\begin{equation}
{\rm BRDF}_{\rm bs}\leq\frac{\epsilon_{\rm max}^2}{2\pi}\left(\frac{R}{\lambda\alpha}\right)^2\left[\ln\left(\frac{\theta_{\rm max}}{\theta_{\rm min}}\right)\right]^{-1},
\qquad n=2.
\label{eq:forward_requirement}
\end{equation}
In contrast with narrow-angle scattering, the tube radius enters explicitly: increasing $R$ reduces the coupling as $R^{-2}$ for a fixed angular range. This makes the dimensions of connecting tubes and the optical quality of their baffles directly relevant to the stray-light requirement.

\subsection{Wide-angle scattering}
\label{sec:wide_scattering}

The final regime corresponds to secondary surfaces located at comparatively large angles from the nominal beam direction. These can include chamber walls, optical table components, structural elements, and dedicated baffles close to the emitting optic. The geometries can vary substantially from one optical chamber to another, so we retain the distance $r$, scattering angle $\theta$, angle $\theta_\perp$ relative to the surface normal, and illuminated area $A_b$ explicitly.

Because the scattering angle can now be large, the cosine factor of the BRDF of the emitting optic in Eq.~\eqref{eq:brdf} must be retained. For the secondary surface we use the conservative bound of Eq.~\eqref{eq:brdf_conservative_bound}, since its detailed orientation is generally layout dependent. For a single backscattering surface, Eq.~\eqref{eq:recoupling} then gives the conservative estimate
\begin{equation}
\epsilon^2=\frac{\lambda^2\alpha^2A_b}{r^4\theta^{2n}}{\rm BRDF}_{\rm bs}\cos^2(\theta_\perp).
\label{eq:wide_general}
\end{equation}
The corresponding requirement can be written as
\begin{equation}
{\rm BRDF}_{\rm bs}\leq\frac{\epsilon_{\rm max}^2r^4\theta^{2n}}{\lambda^2\alpha^2A_b\cos^2(\theta_\perp)}.
\label{eq:wide_requirement}
\end{equation}

The scaling of Eq.~\eqref{eq:wide_compact} makes the competition between distance and scattering angle explicit. Nearby surfaces are particularly important because of the $r^{-4}$ dependence, while surfaces located farther from the main propagation direction receive substantially less light because of the $\theta^{-2n}$ dependence of the main optic BRDF. The most problematic wide-angle surfaces are therefore generally those that combine a relatively short distance from the optic with a sufficiently small scattering angle. Their orientation also matters both through the projected scattering probability represented by $\cos^2(\theta_\perp)$ and through the BRDF of the surface itself.

Equation~\eqref{eq:wide_requirement} provides a direct way to evaluate individual surfaces once the optical layout is known. 

\subsection{Collections of planar baffles}
\label{sec:planar_baffles}

A collection of planar baffles can be used to prevent wide-angle stray light from reaching chamber walls or other uncontrolled surfaces. The individual baffles generally cover different angular regions, have different distances from the emitting optic, and may be tilted to prevent specular reflection back towards the optical path. We therefore extend the single surface expression of Eq.~\eqref{eq:wide_general} by integrating over the solid angle subtended by each baffle.

For the reference $n=2$ BRDF model, we assume that the contributions from the different scattering paths add incoherently. The relative optical phases of these paths are uncontrolled at the sub-wavelength level and depend on microscopic surface and geometrical details, as well as slowly varying environmental conditions. Averaging over these unknown phases therefore removes the cross terms in the recoupled power, so that the total contribution can be written as
\begin{equation}
\epsilon^2=\sum_i\epsilon_i^2,
\label{eq:baffle_sum}
\end{equation}
with
\begin{equation}
\epsilon_i^2=\int_{\Delta\Omega_i}{\rm BRDF}_i(\theta,\phi)\frac{\lambda^2}{r_i^2(\theta,\phi)}\frac{\alpha^2}{\theta^4}\cos^2(\theta_\perp)\td\Omega.
\label{eq:baffle_integral}
\end{equation}
Here $\Delta\Omega_i$ is the solid angle covered by the $i$th baffle and $r_i(\theta,\phi)$ is the distance between the emitting optic and each point on its surface.

To obtain an analytical upper limit, we make a set of conservative approximations. We describe each baffle by the angular intervals $\theta_{{\rm min},i}<\theta<\theta_{{\rm max},i}$ and $\phi_{{\rm min},i}<\phi<\phi_{{\rm max},i}$, evaluate the BRDF and $\cos^2(\theta_\perp)$ factors at their maximum value over the surface, and replace $r_i(\theta,\phi)$ by the minimum distance $r_{{\rm min},i}$. We also restrict the calculation to the portion of the baffle satisfying $\theta_\perp<\pi/2$, since the opposite side of the surface is not directly illuminated.

Let $\Theta_i$ denote the angle between the normal to the $i$th baffle and the nominal beam direction. The incidence angle on the baffle is then characterized by $\Theta_i-\theta$. Defining $\Delta\phi_i=\phi_{{\rm max},i}-\phi_{{\rm min},i}$, Eq.~\eqref{eq:baffle_integral} can be bounded by
\begin{equation}
\epsilon_i^2\lesssim\max\left[{\rm BRDF}_i(\Theta_i-\theta)\cos^2(\theta_\perp)\right]\frac{\alpha^2\lambda^2\Delta\phi_i}{r_{{\rm min},i}^2}\int_{\theta_{{\rm min},i}}^{\theta_{{\rm max},i}}\frac{\sin\theta}{\theta^4}\td\theta.
\label{eq:baffle_bound_integral}
\end{equation}
Using $\sin\theta<\theta$ provides a simple upper limit,
\begin{equation}
\epsilon_i^2\lesssim\frac{\alpha^2\lambda^2\Delta\phi_i}{2r_{{\rm min},i}^2}\left(\frac{1}{\theta_{{\rm min},i}^2}-\frac{1}{\theta_{{\rm max},i}^2}\right)\max\left[{\rm BRDF}_i(\Theta_i-\theta)\cos^2(\theta_\perp)\right].
\label{eq:baffle_bound}
\end{equation}

A useful configuration is one in which the baffle is tilted such that $\Theta_i>\theta_{{\rm max},i}$, so that the normal to the surface lies outside the angular interval subtended by the baffle. If its BRDF does not increase with incidence angle, the BRDF factor is bounded by its value at the smallest incidence angle,
\begin{equation}
\Theta_{{\rm min},i}\equiv\Theta_i-\theta_{{\rm max},i},
\label{eq:theta_min_baffle}
\end{equation}
while the geometrical factor is independently bounded by its maximum value, $\cos^2(\theta_{\perp{\rm min},i})$, where $\theta_{\perp{\rm min},i}$ is the minimum value of $\theta_\perp$ over the illuminated portion of the baffle. Taking these two extrema separately provides a conservative upper bound, even if they do not occur at the same point on the surface, and Eq.~\eqref{eq:baffle_bound} becomes
\begin{equation}
\epsilon_i^2\lesssim\frac{\alpha^2\lambda^2\Delta\phi_i}{2r_{{\rm min},i}^2}\left(\frac{1}{\theta_{{\rm min},i}^2}-\frac{1}{\theta_{{\rm max},i}^2}\right){\rm BRDF}_i(\Theta_{{\rm min},i})\cos^2(\theta_{\perp{\rm min},i}).
\label{eq:baffle_tilted}
\end{equation}

The angular expression can also be rewritten in terms of the physical dimensions of the baffle. Let $w_i$ and $h_i$ denote its width and height, with $w_i=\Delta\phi_i\sin(\theta_{{\rm max},i})r_{{\rm min},i}$ and $h_i\cos(\Theta_{{\rm min},i})$ its projected height. Defining $\bar{\theta}_i=(\theta_{{\rm max},i}+\theta_{{\rm min},i})/2$, we obtain
\begin{equation}
\epsilon_i^2\lesssim\frac{\alpha^2\lambda^2w_ih_i\cos(\Theta_{{\rm min},i})}{r_{{\rm min},i}^4\sin(\theta_{{\rm max},i})}\left(\frac{\bar{\theta}_i}{\theta_{{\rm min},i}^2\theta_{{\rm max},i}^2}\right){\rm BRDF}_i(\Theta_{{\rm min},i})\cos^2(\theta_{\perp{\rm min},i}).
\label{eq:baffle_physical}
\end{equation}
The requirement for the complete baffle system is therefore obtained by summing Eq.~\eqref{eq:baffle_physical} over all illuminated surfaces and imposing
\begin{equation}
\sum_i\epsilon_i^2\leq\epsilon_{\rm max}^2.
\label{eq:baffle_requirement}
\end{equation}

This form makes the main design dependencies explicit. Baffles close to the optic are strongly weighted by the $r_{{\rm min},i}^{-4}$ dependence, while baffles extending towards smaller scattering angles are enhanced by the angular factor involving $\theta_{{\rm min},i}$ and $\theta_{{\rm max},i}$. Tilting the surfaces away from the incident scattered light both reduces their projected area and allows the relevant BRDF to be evaluated farther from the specular direction. Consequently, the total requirement does not need to be distributed equally among all baffles: the geometry naturally identifies which surfaces dominate the recoupling and therefore require the greatest control of their scattering properties.

\section{Ghost beams and beam-dump requirements}
\label{sec:ghost_beams}

A second class of stray-light paths in the core optics is produced by ghost beams. Unlike the diffuse scattering paths considered in Sec.~\ref{sec:analytical_scattering}, the light reaching a beam dump follows a deterministic parasitic optical path, typically generated by residual reflections or transmissions at optical surfaces. These beams should be intercepted by dedicated beam dumps before they illuminate uncontrolled surfaces. However, interception alone does not eliminate the coupling as light reflected or scattered by the dump can propagate back along the reverse ghost path and recouple with the main interferometer field. The relevant beam-dump requirement therefore depends not only on its total reflected power, but also on the angular distribution of that reflected light and on the dump geometry.

Let $P$ be the main beam power incident on the optic from which the ghost path originates and $P_{\rm gh}$ the power carried by the ghost beam. We define the coupling factor as
\begin{equation}
\Gamma_{\rm gh}\equiv\frac{P_{\rm gh}}{P}.
\label{eq:ghost_fraction}
\end{equation}
In contrast with Eq.~\eqref{eq:recoupling}, $\Gamma_{\rm gh}$ is determined by the specular parasitic path rather than by diffuse scattering from the emitting optic. In general,
\begin{equation}
\Gamma_{\rm gh}=\left(\prod_i R_i\right)\left(\prod_j T_j\right)\eta_{\rm path},
\label{eq:ghost_path}
\end{equation}
where $R_i$ and $T_j$ are the power reflectances and transmittances encountered by the ghost beam and $\eta_{\rm path}$ accounts for clipping or other geometrical losses.

We characterize the response of the beam dump by an effective differential return coefficient, denoted ${\rm BRDF}_{\rm bd}^{\rm eff}$ for convenience and defined in the direction of the reverse ghost path. Unlike the microscopic surface BRDF of Eq.~\eqref{eq:brdf}, this is a quantity defined directly as returned power per unit solid angle and therefore already incorporates projection factors, the dump geometry, and multiple reflections inside it. It should consequently be interpreted as an effective BRDF rather than a physical BRDF of any individual surface. If $\delta\Omega_{\rm gh}$ is the solid angle occupied by the reverse ghost mode, the fraction of dump-incident power returned towards that mode is
\begin{equation}
\frac{\delta P_{\rm rev}}{P_{\rm gh}}\simeq{\rm BRDF}_{\rm bd}^{\rm eff}\delta\Omega_{\rm gh}.
\label{eq:ghost_reverse_scattering}
\end{equation}

For a diffraction limited fundamental Gaussian ghost mode with waist radius $w_{\rm gh}$,
\begin{equation}
\theta_{\rm gh}\simeq\frac{\lambda}{\pi w_{\rm gh}},
\label{eq:ghost_divergence}
\end{equation}
and therefore
\begin{equation}
\delta\Omega_{\rm gh}\simeq\pi\theta_{\rm gh}^2=\frac{\lambda^2}{\pi w_{\rm gh}^2}.
\label{eq:ghost_solid_angle}
\end{equation}

By reciprocity, light returned into this mode traverses the same parasitic optical path and couples back to the main field with the same factor $\Gamma_{\rm gh}$. The fraction of main-beam power completing the full path is consequently
\begin{equation}
\epsilon_{\rm gh}^2=\Gamma_{\rm gh}^2{\rm BRDF}_{\rm bd}^{\rm eff}\delta\Omega_{\rm gh}=\frac{\lambda^2}{\pi w_{\rm gh}^2}\Gamma_{\rm gh}^2{\rm BRDF}_{\rm bd}^{\rm eff}.
\label{eq:ghost_recoupling}
\end{equation}
Comparison with the maximum coupling derived in Sec.~\ref{sec:maximum_coupling} gives
\begin{equation}
{\rm BRDF}_{\rm bd}^{\rm eff}\leq\frac{\pi w_{\rm gh}^2}{\lambda^2}\frac{\epsilon_{\rm max}^2}{\Gamma_{\rm gh}^2}.
\label{eq:ghost_requirement}
\end{equation}
The squared dependence on $\Gamma_{\rm gh}$ strongly suppresses weak higher order ghost paths. The most stringent dump requirements are therefore generally associated with a relatively small number of the most powerful low order ghosts.

\begin{figure}[htbp]
    \centering
    \includegraphics[width=1.0\linewidth]{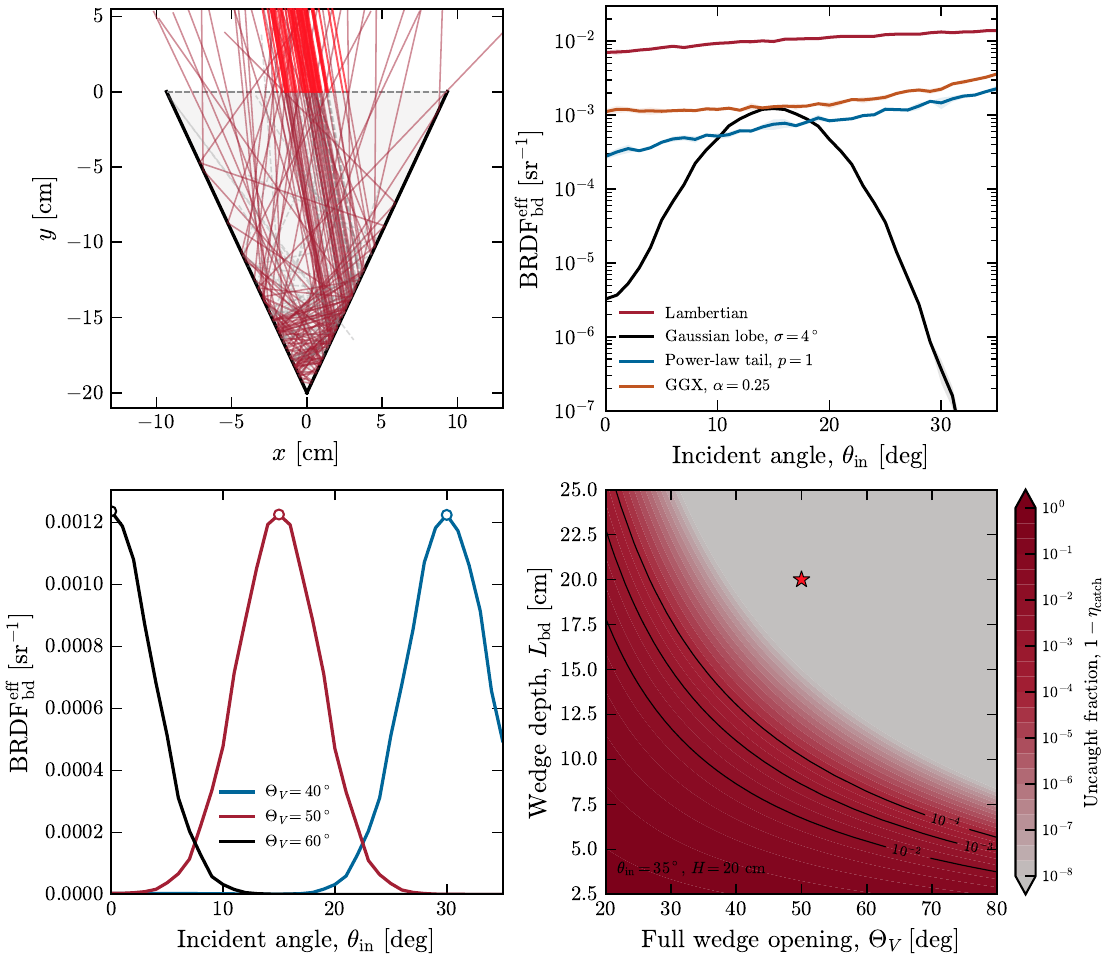}
    \caption{
    \justifying
    Three-dimensional Monte Carlo study of a V-shaped ghost-beam dump. Top left: top-view projection of ray trajectories. Top right: effective BRDF in the reverse ghost direction as a function of incidence angle $\theta_{\rm in}$. Bottom left: dependence on the full wedge opening angle $\Theta_V$; open circles indicate the predicted three-bounce retroreflection angles. Bottom right: fraction of a finite Gaussian beam that is not intercepted as a function of wedge opening and depth for $H_{\rm bd}=20$~cm and $\theta_{\rm in}=35^\circ$. The star denotes the reference geometry for the beam-capture study; other panels scan or use different incidence angles as stated in the text.}
    \label{fig:ghost_dump_mc}
\end{figure}

Equation~\eqref{eq:ghost_requirement} treats ${\rm BRDF}_{\rm bd}^{\rm eff}$ as an input, but the effective return of a real beam dump depends strongly on its geometry. To illustrate this dependence we performed a 3D Monte Carlo ray tracing calculation of a simple V-shaped dump, shown in Fig.~\ref{fig:ghost_dump_mc}. The reference geometry has a wedge full opening angle $\Theta_V=50^\circ$, depth $L_{\rm bd}=20$~cm, height $H_{\rm bd}=20$~cm, and is illuminated by a centered Gaussian beam with local transverse radius $w_{\rm bd}=2$~cm at the dump. We denote by $\theta_{\rm in}$ the angle between the incident beam direction and the symmetry axis of the V-shaped dump in the horizontal plane. At each wall interaction the surviving reflected power is multiplied by a common hemispherical reflected fraction $R_{\rm bd}=0.05$, while its direction is sampled from one of four representative angular scattering models. The resulting ${\rm BRDF}_{\rm bd}^{\rm eff}$ is evaluated as the local returned power per unit solid angle around the exact reverse ghost direction. Details of the Monte Carlo procedure and scattering models are given in App.~\ref{app:ghost_dump_mc}.

The upper-right panel of Fig.~\ref{fig:ghost_dump_mc} demonstrates that surfaces with the same total reflected power can nevertheless differ by more than an order of magnitude in their effective backscattering. Over the incidence angles considered, the diffuse Lambertian model reaches ${\rm BRDF}_{\rm bd}^{\rm eff}\simeq1.4\times10^{-2}$~sr$^{-1}$, whereas the representative GGX and power-law models remain below approximately $3.6\times10^{-3}$ and $2.3\times10^{-3}$~sr$^{-1}$, respectively. The quasi-specular Gaussian model generally produces a much smaller return, except for a pronounced geometrical resonance. For the reference GGX model at $\theta_{\rm in}=10^\circ$, the calculation gives ${\rm BRDF}_{\rm bd}^{\rm eff}\simeq1.2\times10^{-3}$~sr$^{-1}$. These values should not be interpreted as predictions for particular materials, since the four models are intentionally compared at the same $R_{\rm bd}$. They instead demonstrate that controlling the angular scattering distribution can be as important as reducing the total reflectivity.

The resonance of the quasi specular case illustrates a second important design consideration. A V-shaped dump can itself form a multibounce retroreflector for particular combinations of incidence and opening angle. For the three-reflection path of the symmetric geometry considered here, exact geometrical return occurs at
\begin{equation}
\theta_{\rm retro}^{(3)}=90^\circ-\frac{3}{2}\Theta_V.
\label{eq:ghost_retroreflection}
\end{equation}
The Monte Carlo peaks occur at $30^\circ$, $15^\circ$, and $0^\circ$ for $\Theta_V=40^\circ$, $50^\circ$, and $60^\circ$, respectively, in agreement with Eq.~\eqref{eq:ghost_retroreflection}. A highly specular surface is therefore not automatically advantageous as, although it can strongly suppress diffuse return over most incidence angles, an unfavorable dump geometry can redirect a significant fraction of the surviving light towards the incident path. The opening angle should consequently be chosen together with the expected ghost beam incidence angles and the scattering profile of the absorbing surface.

The finite dimensions of the dump introduce a complementary constraint. The lower-right panel of Fig.~\ref{fig:ghost_dump_mc} shows the fraction of the incident Gaussian beam that misses the wedge as a function of $\Theta_V$ and $L$. For the representative $20$~cm deep geometry, essentially the complete $w_{\rm bd}=2$~cm beam is intercepted even at $\theta_{\rm in}=35^\circ$, with the calculated uncaught fraction below $3\times10^{-14}$. Once the dump is sufficiently large to intercept the beam, increasing its overall scale does not substantially change the angular return of a geometrically similar V-shaped configuration. The opening angle and surface-scattering law instead control the multiple-bounce trajectories. The appropriate dimensions can therefore first be chosen to guarantee sufficient beam capture, after which the angular geometry can be optimized against retroreflection.

The optical coating provides an additional and potentially powerful degree of freedom. Each additional wall interaction introduces another factor of the reflected fraction $R_{\rm bd}$. In the Monte Carlo calculation the reference GGX configuration scales approximately as ${\rm BRDF}_{\rm bd}^{\rm eff}\propto R_{\rm bd}^{1.15}$, indicating that its return is dominated by low-order paths, while the quasi-specular three-bounce resonance follows ${\rm BRDF}_{\rm bd}^{\rm eff}\propto R_{\rm bd}^3$ as expected. Reducing the reflectivity can therefore suppress problematic multi-bounce paths much more strongly than suggested by a single-reflection BRDF alone. This provides a direct motivation for low-reflectivity or absorbing coatings in addition to a favorable dump geometry.

Finally, it is useful to consider the hierarchy of ghost beams generated by a transmissive optic. Let surfaces 1 and 2 denote its front and back surfaces. Representative first-, second-, and third-order paths scale approximately as
\begin{equation}
\Gamma_{\rm gh}^{(1)}\simeq T_1T_2,\qquad \Gamma_{\rm gh}^{(2)}\simeq T_1R_2R_1T_2,\qquad \Gamma_{\rm gh}^{(3)}\simeq T_1R_2T_1.
\label{eq:ghost_orders}
\end{equation}
For a highly reflective front surface and an antireflection-coated back surface, $R_1\simeq1$ and $T_2\simeq1$, giving
\begin{equation}
\Gamma_{\rm gh}^{(1)}\sim T_1,\qquad \Gamma_{\rm gh}^{(2)}\sim T_1R_2,\qquad \Gamma_{\rm gh}^{(3)}\sim T_1^2R_2.
\label{eq:ghost_hierarchy}
\end{equation}
Combined with the $\Gamma_{\rm gh}^2$ dependence of Eq.~\eqref{eq:ghost_recoupling}, this means that the beam-dump design will normally be driven by the strongest low-order ghost paths. For each of these paths, Eq.~\eqref{eq:ghost_requirement} provides the required effective BRDF, while Fig.~\ref{fig:ghost_dump_mc} illustrates why satisfying that requirement should be treated as a combined problem of surface coating, dump geometry, and ghost-beam incidence rather than as a material property alone.

\section{Application to Cosmic Explorer}
\label{sec:CE}

We now apply the analytical framework developed in the previous sections to a representative CE~\cite{CE1,CE2,CE3} configuration. The purpose is not to define a final stray light design for CE's core optics, since the detailed optical layout and its mechanical interfaces are still evolving, but to translate the expected detector sensitivity into approximate requirements that can be used to guide that design. Unless stated otherwise, we use the parameters listed in Tab.~\ref{tab:CE_parameters}. The optical scattering model corresponds to $\alpha=10^{-6}$ and $n=2$, representative of good quality optical surfaces over the angular range considered here~\cite{Thorne89,Romero-Rodriguez:2022mje}.

\begin{table}[htbp]
\centering
\caption{\justifying Representative parameters used for the Cosmic Explorer application. These values are intended as a reference configuration rather than binding design parameters.}
\label{tab:CE_parameters}
\begin{tabular}{lll}
\hline
Parameter & Description & Value \\
\hline
$\alpha$ & BRDF coefficient of Eq.~\eqref{eq:brdf_powerlaw} & $10^{-6}$ \\
$n$ & BRDF power-law index & $2$ \\
$\lambda$ & Laser wavelength & $1064$ nm \\
$R$ & Inner radius of connecting tube & $0.5$ m \\
$R_t$ & Characteristic radius of emitting-optic chamber & $1$ m \\
$R_b$ & Clear radius of tube baffles & $0.4$ m \\
$R_{ts}$ & Characteristic radius of target-optic chamber & $1$ m \\
$R_m$ & Radius of target optic & $0.15$ m \\
$L_s$ & Representative separation between core optics & $50$ m \\
$L_a$ & Arm length & $40$ km \\
$\mathcal{F}_a$ & Arm cavity finesse & $450$ \\
\hline
\end{tabular}
\end{table}

The first step is to determine the maximum amount of light that can recouple to the interferometer. Following Eq.~\eqref{eq:seismic_model}, we describe the displacement ASD of an unisolated backscattering surface by $x_{\rm bs}=A/f^2$. A value $A\simeq5\times10^{-8}$~m~Hz$^{3/2}$ provides a representative description of the median ground motion observed at LIGO Livingston, while motion approaching the upper part of the observed distribution can be approximately represented by $A\simeq5\times10^{-7}$~m~Hz$^{3/2}$~\cite{Andres-Carcasona:2023qom}. We use the latter as the conservative reference for surfaces attached directly to the vacuum system and scan several values of $A$ to illustrate the effect of the mechanical environment.

\begin{figure}[htbp]
    \centering
    \includegraphics[width=1.0\linewidth]{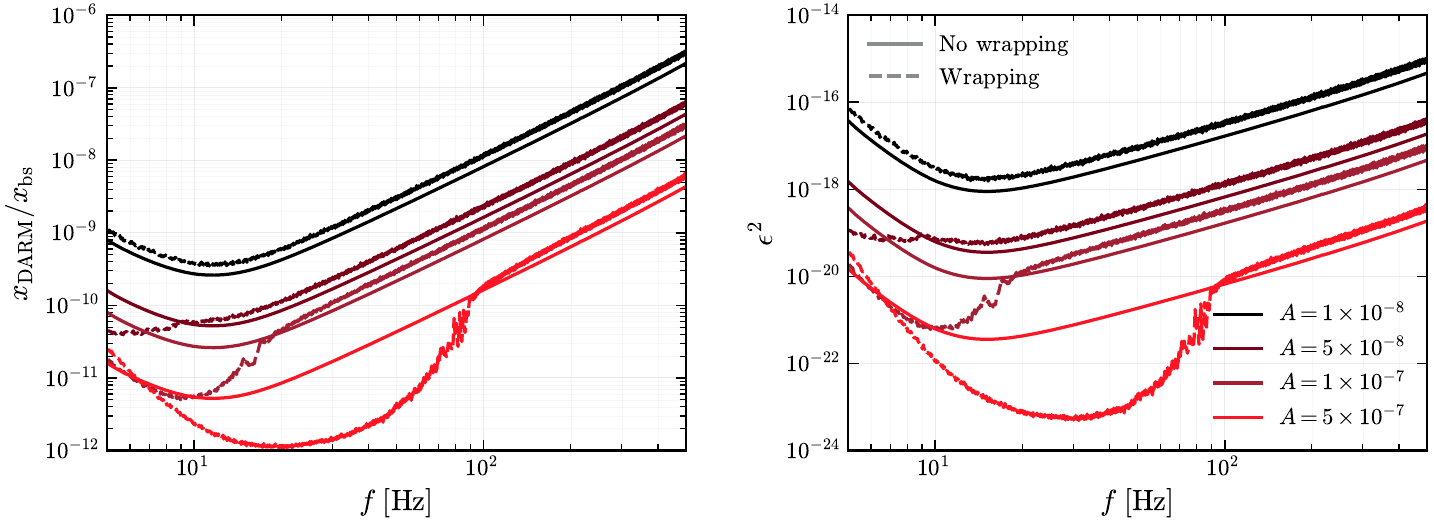}
    \caption{\justifying Maximum stray-light recoupling for representative CE surface-motion spectra. Left: ratio between the CE DARM displacement target and the backscattering surface motion for different values of $A$, with and without phase wrapping. Right: corresponding maximum recoupled power fraction obtained from Eq.~\eqref{eq:epsilon_max}.}
    \label{fig:CE_max_coupling}
\end{figure}

Substituting the CE target $x_{\rm DARM}$ and the displacement models (see Ref.~\cite{Evans:2021gyd} and, in particular, we use the curves of Ref.~\cite{CE_sens}) into Eq.~\eqref{eq:epsilon_max} gives the maximum recoupling shown in Fig.~\ref{fig:CE_max_coupling}. The calculation is shown both using the original displacement spectrum and using the effective spectrum obtained from the phase-wrapping procedure of Sec.~\ref{sec:motion_wrapping}. For sufficiently small motion the two descriptions converge, whereas for the larger displacement levels the phase-wrapped spectrum contains substantially more power in the observing band and produces a correspondingly tighter recoupling requirement. In the conservative $A=5\times10^{-7}$~m~Hz$^{3/2}$ case, the most restrictive value is approximately
\begin{equation}
\epsilon_{\rm max}^2\simeq1.6\times10^{-24}.
\label{eq:CE_epsilon_baseline}
\end{equation}
This value corresponds to taking the full CE DARM target as the upper-level scattered light allowance and any more restrictive subsystem noise allocation can be incorporated directly through the $\eta_{\rm SL}^2$ scaling described in Sec.~\ref{sec:maximum_coupling}.

The value in Eq.~\eqref{eq:CE_epsilon_baseline} corresponds to a surface experiencing the conservative ground motion spectrum and significant phase wrapping. Other mechanical configurations can be incorporated without redefining all of the optical requirements. We introduce a phase wrapping penalty $F_{\rm wrap}$, normalized such that $F_{\rm wrap}=1$ when phase wrapping is negligible and $F_{\rm wrap}\simeq200$ for the conservative unisolated reference case. For an isolated surface, $F_{\rm wrap}$ is evaluated from the residual motion after applying the isolation factor $H_{\rm iso}$ of Eq.~\eqref{eq:isolation}, so the two quantities should not be interpreted as independent corrections. With this convention, the maximum allowed coupling can be written approximately as
\begin{equation}
\epsilon_{\rm max}^2\simeq1.6\times10^{-24}\left(\frac{200}{F_{\rm wrap}}\right)H_{\rm iso}^2\equiv1.6\times10^{-24}F_{\rm atten},
\label{eq:CE_epsilon_fatten}
\end{equation}
where
\begin{equation}
F_{\rm atten}\equiv\left(\frac{200}{F_{\rm wrap}}\right)H_{\rm iso}^2.
\label{eq:Fatten}
\end{equation}
This factor conveniently combines the mechanical and nonlinear contributions to the requirement. Representative configurations range from $F_{\rm atten}\simeq1$ for an unisolated surface with substantial phase wrapping, to $F_{\rm atten}\simeq6$ for mild isolation while phase wrapping remains relevant, $F_{\rm atten}\simeq6\times10^3$ for a well-isolated surface for which phase wrapping is avoided, and $F_{\rm atten}\simeq6\times10^5$ for a strongly isolated surface. These values should be interpreted as representative consistent mechanical regimes rather than universal properties of particular detector components. In particular, arbitrary combinations of $H_{\rm iso}$ and $F_{\rm wrap}$ are not implied.

We can now substitute Eq.~\eqref{eq:CE_epsilon_fatten} into the analytical expressions of Sec.~\ref{sec:analytical_scattering}. For narrow-angle scattering, Eq.~\eqref{eq:narrow_requirement} becomes
\begin{equation}
{\rm BRDF}_{\rm bs}\left(\frac{A_b}{0.5~{\rm m}^2}\right)\left(\frac{0.15~{\rm m}}{R_m}\right)^4\lesssim1.4\times10^{-3}F_{\rm atten}~{\rm sr}^{-1}.
\label{eq:CE_narrow_requirement}
\end{equation}
This is the most restrictive of the three single-surface estimates for the reference geometry. For example, a surface with ${\rm BRDF}_{\rm bs}=10^{-2}$~sr$^{-1}$ and $A_b=0.5$~m$^2$ surrounding a $R_m=0.15$~m optic would require $F_{\rm atten}\gtrsim7$. This is consistent with narrow angle baffles being preferentially mounted on an isolated structure or mechanically referenced to the target optic so as to minimize the relative motion that changes the scattered light path length, rather than being attached directly to the vacuum chamber.

For forward scattering, the representative geometry of Table~\ref{tab:CE_parameters} gives
\begin{equation}
\theta_{\rm min}\simeq\frac{R_b}{L_s-R_{ts}}\simeq0.47^\circ,\qquad \theta_{\rm max}\simeq\arctan\left(\frac{R}{R_t}\right)\simeq26.6^\circ,
\end{equation}
and therefore
\begin{equation}
\ln\left(\frac{\theta_{\rm max}}{\theta_{\rm min}}\right)\simeq4.04.
\end{equation}
Equation~\eqref{eq:forward_requirement} then gives
\begin{equation}
{\rm BRDF}_{\rm bs}\left(\frac{0.5~{\rm m}}{R}\right)^2\left[\frac{\ln(\theta_{\rm max}/\theta_{\rm min})}{4.04}\right]\lesssim1.4\times10^{-2}F_{\rm atten}~{\rm sr}^{-1}.
\label{eq:CE_forward_requirement}
\end{equation}
Unlike the narrow-angle case, a surface with a BRDF of order $10^{-2}$~sr$^{-1}$ already satisfies the representative requirement even for $F_{\rm atten}\simeq1$. This suggests that approximately axisymmetric annular tube baffles represented by this model need not necessarily be mechanically isolated provided that their backscattering properties can be maintained at or below this level.

For wide-angle scattering, taking as a representative example $r=0.5$~m, $A_b=0.5$~m$^2$, and $\theta\simeq\theta_\perp\simeq30^\circ$, Eq.~\eqref{eq:wide_requirement} becomes
\begin{equation}
{\rm BRDF}_{\rm bs}\left(\frac{0.5~{\rm m}}{r}\right)^4\left(\frac{A_b}{0.5~{\rm m}^2}\right)\left[\frac{\cos^2(\theta_\perp)/\theta^4}{12}\right]\lesssim1.7\times10^{-2}F_{\rm atten}~{\rm sr}^{-1},
\label{eq:CE_wide_requirement}
\end{equation}
where the angles appearing in the geometrical factor are expressed in radians. As for the forward-scattering case, a BRDF of order $10^{-2}$~sr$^{-1}$ is sufficient for the representative geometry even without isolation. The strong $r^{-4}$ dependence nevertheless means that individual surfaces very close to an optic should be checked explicitly, particularly if they extend towards smaller scattering angles.

For a collection of planar wide angle baffles, the result of Sec.~\ref{sec:planar_baffles} can be cast into a useful normalized form,
\begin{equation}
\sum_i {\rm BRDF}_{{\rm bs},i}
\left(\frac{A_i\cos\Theta_{{\rm min},i}}{4\pi r_{{\rm min},i}^2}\right)
\left(\frac{0.5~{\rm m}}{r_{{\rm min},i}}\right)^2
\left[\frac{\cos(\theta_{\perp{\rm min},i})}{\theta_{{\rm min},i}\theta_{{\rm max},i}}\right]^2
\left[\frac{\bar{\theta}_i}{\sin(\theta_{{\rm max},i})}\right]
\lesssim3\times10^{-2}F_{\rm atten}~{\rm sr}^{-1},
\label{eq:CE_planar_requirement}
\end{equation}
where $A_i=w_ih_i$. The first parenthesis represents the fraction of the spherical area at $r_{{\rm min},i}$ covered by the projected baffle surface, while the remaining terms encode its distance and angular position. This form is particularly useful during the design of an optical chamber because the contribution of each proposed baffle can be evaluated independently and the total compared directly with the requirement. It also makes clear that the closest baffles extending towards the smallest values of $\theta$ will generally dominate the sum.

The corresponding requirement for ghost-beam dumps follows from Eq.~\eqref{eq:ghost_requirement} and for the CE reference wavelength it can be written as
\begin{equation}
{\rm BRDF}_{\rm bd}^{\rm eff}\left(\frac{\Gamma_{\rm gh}}{1~{\rm ppm}}\right)^2\left(\frac{1~{\rm cm}}{w_{\rm gh}}\right)^2\lesssim5\times10^{-4}F_{\rm atten}~{\rm sr}^{-1}.
\label{eq:CE_ghost_requirement}
\end{equation}
The strong $\Gamma_{\rm gh}^2$ dependence makes the requirement particularly sensitive to the power of the lowest-order ghost beams. As an illustration, consider the coating values discussed in Sec.~\ref{sec:ghost_beams}, $T_1\simeq10^{-5}$ and $R_2\simeq10^{-4}$, with $R_1\simeq T_2\simeq1$. These give $\Gamma_{\rm gh}^{(1)}\sim10^{-5}$, $\Gamma_{\rm gh}^{(2)}\sim10^{-9}$, and $\Gamma_{\rm gh}^{(3)}\sim10^{-14}$. The second and third order paths are therefore strongly suppressed, while the first order ghost corresponds to approximately $10$~ppm and can set a meaningful beam dump requirement.

Taking $w_{\rm gh}=2$~cm as a representative Gaussian mode waist radius, a $10$~ppm first order ghost beam requires
\begin{equation}
{\rm BRDF}_{\rm bd}^{\rm eff}\lesssim2\times10^{-5}F_{\rm atten}~{\rm sr}^{-1}.
\label{eq:CE_ghost_first_order}
\end{equation}
This allows the Monte Carlo results of Fig.~\ref{fig:ghost_dump_mc} to be compared directly with the CE requirement. The reference GGX configuration gives ${\rm BRDF}_{\rm bd}^{\rm eff}\simeq1.2\times10^{-3}$~sr$^{-1}$ at $\theta_{\rm in}=10^\circ$, implying that this illustrative $R_{\rm bd}=0.05$ dump would require approximately
\begin{equation}
F_{\rm atten}\gtrsim60
\label{eq:CE_ghost_GGX_fatten}
\end{equation}
for such a first-order ghost. A Lambertian scattering model at the largest value found in the Monte Carlo, ${\rm BRDF}_{\rm bd}^{\rm eff}\simeq1.4\times10^{-2}$~sr$^{-1}$, would instead require $F_{\rm atten}\gtrsim700$. These numbers should not be interpreted as requirements on a specific coating, since the Monte Carlo scattering models were intentionally normalized to the same illustrative $5\%$ power reflectance. They demonstrate, however, that the first order ghost beam can be substantially more demanding than the higher order paths and that a successful dump design may require a combination of lower reflectivity, favorable angular scattering, avoidance of geometrical retroreflection, and mechanical isolation.

Taken together, Eqs.~\eqref{eq:CE_narrow_requirement}, \eqref{eq:CE_forward_requirement}, \eqref{eq:CE_wide_requirement}, \eqref{eq:CE_planar_requirement}, and \eqref{eq:CE_ghost_requirement} provide a first set of quantitative core-optics stray-light requirements for CE and are the ones currently included in the stray light design documentation. For representative scattering surfaces with ${\rm BRDF}_{\rm bs}\sim10^{-2}$~sr$^{-1}$, forward and wide-angle scattering can satisfy the requirement even for surfaces that approximately follow ground motion, whereas narrow-angle surfaces require a modest mechanical isolation or optical coating (or, ideally, both). Ghost beams are qualitatively different: because the strongest low-order paths can carry substantially more power than the diffuse-scattering fields, their beam dumps can require either considerably smaller effective backscattering or stronger mechanical isolation. The detailed balance between these strategies will ultimately depend on the final CE optical layout, the measured BRDFs of candidate surfaces, and the mechanical implementation of the corresponding baffles and beam dumps.

\section{Conclusions}
\label{sec:conclusions}

We have developed an analytical framework to estimate stray light noise from the core optics region of ground based gravitational wave interferometers and translate detector sensitivity into quantitative requirements on the surfaces used to control it. The central quantity is the fraction of optical power $\epsilon^2$ that completes a parasitic scattering path and recouples with the main optical mode. Relating this quantity to the equivalent DARM displacement noise allows the optical problem to be separated from the mechanical motion of the backscattering surface. In particular, mechanical isolation directly relaxes the allowed optical coupling, while sufficiently large motion introduces phase wrapping and can make the requirement substantially more restrictive through the upconversion of low-frequency displacement.

The diffuse scattering problem was divided into narrow-angle, forward, and wide-angle regimes according to the location of the secondary scattering surface. This separation leads to simple analytical expressions that retain the dominant geometrical dependencies. We further generalized the wide-angle calculation to collections of planar baffles, providing an upper limit expression that can be directly evaluated for a proposed chamber geometry and naturally identifies the surfaces that dominate the stray light budget. Ghost beams require a different treatment because the outgoing parasitic field is generated deterministically by optical reflections and transmissions rather than by diffuse scattering. The effective BRDF of the dump is not solely a material property, since multiple reflections and the dump geometry can strongly modify the probability of returning light along the reverse ghost path. Our Monte Carlo study of a V-shaped dump illustrates this explicitly. Surfaces having the same total reflected fraction can produce more than an order of magnitude difference in effective backscattering depending on their angular scattering distribution, and quasi specular surfaces can exhibit multibounce retroreflection resonances for particular combinations of wedge and incidence angles. Conversely, reducing the reflected fraction suppresses multiplebounce paths increasingly strongly with their reflection order. These results emphasize that ghost-beam dumps should be designed by considering surface scattering, absorption, geometry, and expected incidence angles together rather than optimizing any one quantity independently.

We applied the framework to a representative Cosmic Explorer configuration, which has been used to set the current stray light requirements in this area before a mature design is available. These numerical requirements should not be interpreted as final specifications for individual components, since they use representative dimensions, optical properties, and mechanical environments. This is also the regime in which the analytical approach is most useful, as the expressions expose the dominant scalings and allow requirements to be updated immediately as apertures, distances, scattering measurements, isolation performance, or ghost beam paths become better defined. Once a detailed optical configuration is fixed, these estimates can be complemented by comprehensive simulations for the paths identified as most critical.

The main design implication is therefore that core optics stray light control should be treated as a coupled optical and mechanical problem. The scattering properties of a surface cannot be assigned a meaningful requirement independently of its location and motion, and mechanical isolation can in some cases provide more margin than further reductions in BRDF. Conversely, the geometry of baffles and beam dumps can determine whether otherwise acceptable materials produce problematic return paths. The analytical requirements derived here provide a compact way of making these tradeoffs during detector design and constitute a starting point for the detailed stray light control strategy of Cosmic Explorer and other future gravitational wave interferometers.

\ack{This research has been supported by the National Science Foundation (NSF) awards PHY-2309064, PHY-2308793 and PHY-2308794. This document has received a CE DCC number of CE-P2600005.}

\data{All the results can be reproduced following the calculations contained in the manuscript.}

\bibliographystyle{iopart-num}
\bibliography{references}

\appendix
\section{BRDF model for tilted core optics}
\label{app:tilted_brdf}

The power law model of Eq.~\eqref{eq:brdf_powerlaw} is written only in terms of the scattering angle relative to the specular direction. Several core optics, however, operate at non-normal incidence. Here we show that the same parametrization remains a useful approximation for the reference $n=2$ case considered throughout this work.

For an optically smooth surface, Rayleigh--Rice perturbation theory relates the BRDF to the two-dimensional surface-height power spectral density (PSD) through~\cite{stover2012optical}
\begin{equation}
{\rm BRDF}(\theta_i,\theta_o)=\frac{16\pi^2}{\lambda^4}Q\cos(\theta_i)\cos(\theta_o)\,
{\rm PSD}_{2{\rm D}}\left(\frac{|\sin(\theta_o)-\sin(\theta_i)|}{\lambda}\right),
\label{eq:app_brdf_psd}
\end{equation}
where $\theta_i$ and $\theta_o$ are the incident and outgoing angles relative to the surface normal and $Q$ is the polarization-dependent reflection factor. Here we consider scattering in the plane of incidence. For an isotropic surface the two-dimensional PSD depends only on the magnitude of the spatial-frequency vector, so it can be written as a function of the single scalar spatial frequency appearing in Eq.~\eqref{eq:app_brdf_psd}. We absorb the variation of $Q$ over the angular range of interest into the overall normalization.

Assuming a power-law surface spectrum
\begin{equation}
{\rm PSD}_{2{\rm D}}(\xi)=A\xi^{-m},
\end{equation}
Eq.~\eqref{eq:app_brdf_psd} becomes
\begin{equation}
{\rm BRDF}(\theta_i,\theta_o)=K\frac{\cos(\theta_i)\cos(\theta_o)}{|\sin(\theta_o)-\sin(\theta_i)|^m},
\label{eq:app_brdf_tilt_general}
\end{equation}
where $K=16\pi^2QA\lambda^{m-4}$. Defining the scattering angle $\theta$ relative to the specular direction as in Fig.~\ref{fig:scattering_regions}, one has $\theta_o=\theta_i-\theta$, and therefore
\begin{equation}
{\rm BRDF}(\theta;\theta_i)=K\frac{\cos(\theta_i)\cos(\theta_i-\theta)}{|\sin(\theta_i-\theta)-\sin(\theta_i)|^m}.
\label{eq:app_brdf_tilt}
\end{equation}

At normal incidence, $\theta_i=0$, this reduces to
\begin{equation}
{\rm BRDF}(\theta;0)=K\frac{\cos\theta}{\sin^m\theta},
\end{equation}
which approaches $K/\theta^m$ for sufficiently small scattering angles. We can therefore identify $K=\alpha$ and $m=n$ with the parametrization used in Eq.~\eqref{eq:brdf_powerlaw}.

Of particular interest is the $n=2$ case employed for the CE reference calculation. Expanding Eq.~\eqref{eq:app_brdf_tilt} for small $\theta$ gives
\begin{equation}
{\rm BRDF}(\theta;\theta_i)=\frac{\alpha}{\theta^2}\left[1-\left(\frac{1}{6}+\frac{\tan^2\theta_i}{4}\right)\theta^2+\mathcal{O}(\theta^3)\right].
\label{eq:app_brdf_tilt_expansion}
\end{equation}
The leading $\theta^{-2}$ term is therefore independent of the angle of incidence, with the effect of the optic tilt entering only at subleading order. Moreover, the first correction in Eq.~\eqref{eq:app_brdf_tilt_expansion} is negative, such that the pure $\alpha/\theta^2$ model generally provides a conservative approximation over the range where the expansion is applicable.

As a consistency check, the power law BRDF can also be related to the total integrated scattering (TIS) of the optic. Assuming azimuthal symmetry,
\begin{equation}
{\rm TIS}=2\pi\int_{\theta_{\rm min}}^{\pi/2}{\rm BRDF}(\theta)\cos\theta\sin\theta\,\td\theta.
\label{eq:app_tis}
\end{equation}
For Eq.~\eqref{eq:brdf_powerlaw},
\begin{equation}
{\rm TIS}=2\pi\alpha\int_{\theta_{\rm min}}^{\pi/2}\frac{\cos\theta\sin\theta}{\theta^n}\,\td\theta.
\end{equation}
For $n=2$ this has the exact form
\begin{equation}
{\rm TIS}=2\pi\alpha\left[{\rm Ci}(\pi)-{\rm Ci}(2\theta_{\rm min})+\frac{\sin(2\theta_{\rm min})}{2\theta_{\rm min}}\right],
\label{eq:app_tis_exact}
\end{equation}
where ${\rm Ci}$ denotes the cosine integral. For $\theta_{\rm min}\ll1$, it is useful to write
\begin{equation}
{\rm TIS}\simeq K_{\rm TIS}(\theta_{\rm min})\alpha,
\end{equation}
with
\begin{equation}
K_{\rm TIS}(\theta_{\rm min})\simeq24.2-14.47\log_{10}\left(\frac{\theta_{\rm min}}{1^\circ}\right).
\label{eq:app_tis_approx}
\end{equation}
For the reference $\alpha=10^{-6}$, taking $\theta_{\rm min}=0.01^\circ$ as a representative lower boundary of the angular range over which the power-law BRDF is applied gives ${\rm TIS}\simeq53$~ppm from Eq.~\eqref{eq:app_tis_approx}. The power-law model is not intended to describe the specular peak as $\theta\rightarrow0$. The resulting TIS is consistent with the scattering level expected for high quality interferometer core optics and provides a useful consistency check on the adopted BRDF normalization over the angular range of interest.

\section{Mode recoupling of scattered light}
\label{app:mode_recoupling}

The factor $(\lambda/r)^2$ appearing in Eq.~\eqref{eq:recoupling} represents the efficiency with which scattered light returning from a surface at a distance $r$ can couple to a single optical mode. Although this scaling follows from the Thorne--Flanagan formalism~\cite{Thorne89,Flanagan94,Flanagan95,FlanaganThorne95_Diff}, its physical origin can be understood from a simple Gaussian beam argument.

Consider a scatterer located a distance $r$ from an optic carrying a Gaussian mode of wavelength $\lambda$, waist $w_0$, and Rayleigh range
\begin{equation}
z_R=\frac{\pi w_0^2}{\lambda}.
\end{equation}
We estimate the recoupling probability as the product of two factors: the solid angle occupied by the beam as seen from the scatterer and the probability that light propagating within that region overlaps with the Gaussian spatial mode.

At a distance $z$ from the waist, the Gaussian beam radius is
\begin{equation}
w^2(z)=w_0^2\left(1+\frac{z^2}{z_R^2}\right),
\end{equation}
so that the solid angle subtended by the beam at the scatterer is approximately
\begin{equation}
\delta\Omega_{\rm bs}\simeq\frac{\pi w^2(z)}{r^2}=\frac{\lambda z_R}{r^2}\left(1+\frac{z^2}{z_R^2}\right).
\label{eq:app_solid_angle}
\end{equation}

In the far-field limit, $z\gg z_R$, the beam radius becomes
\begin{equation}
w^2(z)\simeq\frac{\lambda z^2}{\pi z_R},
\end{equation}
and Eq.~\eqref{eq:app_solid_angle} reduces to
\begin{equation}
\delta\Omega_{\rm bs}\simeq\frac{\lambda z^2}{z_Rr^2}.
\end{equation}
The waist can then be regarded approximately as a target of area $\pi w_0^2$ located a distance $z$ away. For isotropic scattering into the forward hemisphere, the probability of reaching this target is
\begin{equation}
P_{\rm coup}\simeq\frac{\pi w_0^2}{2\pi z^2}=\frac{\lambda z_R}{2\pi z^2}.
\end{equation}
The product is therefore
\begin{equation}
\delta\Omega_{\rm bs}P_{\rm coup}\simeq\frac{1}{2\pi}\left(\frac{\lambda}{r}\right)^2.
\label{eq:app_farfield_recoupling}
\end{equation}

The same scaling is obtained in the near field by considering the angular acceptance of the Gaussian mode. At the waist, its characteristic divergence is
\begin{equation}
\theta_0=\frac{\lambda}{\pi w_0}.
\end{equation}
For isotropic forward scattering, the fraction of the angular distribution overlapping the Gaussian mode is of order
\begin{equation}
P_{\rm coup}\simeq\frac{\theta_0^2}{2}.
\end{equation}
At the same time,
\begin{equation}
\delta\Omega_{\rm bs}\simeq\frac{\pi w_0^2}{r^2}=\frac{\lambda z_R}{r^2}.
\end{equation}
Using $\theta_0^2=\lambda/(\pi z_R)$ again gives
\begin{equation}
\delta\Omega_{\rm bs}P_{\rm coup}\simeq\frac{1}{2\pi}\left(\frac{\lambda}{r}\right)^2.
\label{eq:app_nearfield_recoupling}
\end{equation}

The numerical prefactor in Eqs.~\eqref{eq:app_farfield_recoupling} and \eqref{eq:app_nearfield_recoupling} depends on the simplified assumption of isotropic forward scattering and should not be identified with the normalization of Eq.~\eqref{eq:recoupling}. The important result is the scaling
\begin{equation}
P_{\rm recoup}\propto\left(\frac{\lambda}{r}\right)^2.
\end{equation}
The cancellation of the beam size and Rayleigh range reflects the diffraction limited phase space occupied by a single spatial mode as increasing the beam area decreases its angular acceptance by the corresponding amount.

\section{Beam dump Monte Carlo and scattering models}
\label{app:ghost_dump_mc}

The calculation used to illustrate the beam-dump requirement of Sec.~\ref{sec:ghost_beams} consists of a 3D Monte Carlo ray trace through a finite V-shaped geometry. The two absorbing surfaces form a symmetric wedge with full opening angle $\Theta_V$, depth $L_{\rm bd}$, and height $H_{\rm bd}$. The wedge is open at its entrance and at the upper and lower boundaries. The center of the incident ghost beam is placed at the same height as the center of the wedge, and its propagation direction is varied within the horizontal plane. Individual rays are sampled from a circular Gaussian intensity distribution with local transverse radius $w_{\rm bd}$ at the dump. The reference calculations use $3\times10^6$ rays, while the parameter scans use two independent realizations of $3.5\times10^5$ rays per configuration. Angular scans are performed in $1^\circ$ steps. 

At each interaction the ray power is multiplied by the hemispherical reflected fraction $R_{\rm bd}$ of the surface. The outgoing direction is then sampled from the normalized angular distribution associated with the chosen BRDF shape. For an isotropic surface, the probability density for an outgoing polar angle $\theta_o$ and azimuth $\phi_o$, at local incidence angle $\theta_i$, is
\begin{equation}
p(\theta_o,\phi_o|\theta_i)=
\frac{
f_r(\theta_i,\theta_o,\phi_o)\cos\theta_o\sin\theta_o
}{
\displaystyle
\int f_r(\theta_i,\theta_o',\phi_o')
\cos\theta_o'\sin\theta_o'\,
\td\theta_o'\td\phi_o'
},
\label{eq:mc_sampling_pdf}
\end{equation}
where $f_r$ specifies the angular shape of the scattering distribution. Its overall normalization therefore does not enter the directional sampling; the total hemispherical reflected fraction is accounted for separately through the factor $R_{\rm bd}$ applied to the ray weight.

For a ray leaving through the entrance we define its angular separation $\psi$ from the exact reverse ghost direction. Since the microscopic BRDF projection factor is already included in the sampling probability of Eq.~\eqref{eq:mc_sampling_pdf}, the effective quantity defined in Sec.~\ref{sec:ghost_beams} is obtained directly from the returned power per unit solid angle. A finite cone estimate is therefore
\begin{equation}
{\rm BRDF}_{\rm bd}^{\rm eff}(\psi_c)=\frac{1}{2\pi(1-\cos\psi_c)}\frac{P(\psi<\psi_c)}{P_{\rm dump}},
\label{eq:mc_brdf_cone}
\end{equation}
where $P_{\rm dump}$ is the power incident on the beam dump. The optical mode appearing in Eq.~\eqref{eq:ghost_solid_angle} occupies a much smaller angular region than can be efficiently sampled by brute force Monte Carlo. We therefore estimate the local power density at $\psi=0$ using a spherical kernel density estimator and verify it against Eq.~\eqref{eq:mc_brdf_cone} for several finite cone sizes. 
%For the reference GGX configuration at $\theta_{\rm in}=10^\circ$, direct estimates using cone half-angles between $1^\circ$ and $4^\circ$ lie between approximately $1.05\times10^{-3}$ and $1.11\times10^{-3}$~sr$^{-1}$, while varying the kernel bandwidth between $1.5^\circ$ and $4^\circ$ gives values between approximately $1.09\times10^{-3}$ and $1.16\times10^{-3}$~sr$^{-1}$. The final value is therefore insensitive to the angular estimator at the level relevant for the design study.

\subsection{Representative scattering models}

The purpose of the surface models used in Fig.~\ref{fig:ghost_dump_mc} is not to assign a specific BRDF to a particular beam dump coating, but to span qualitatively different scattering behaviors while keeping the total hemispherical reflected fraction fixed. We use $R_{\rm bd}=0.05$ for all four cases so that differences in ${\rm BRDF}_{\rm bd}^{\rm eff}$ arise exclusively from the angular redistribution of the reflected power.

Let $\Delta$ denote the angular separation between the outgoing ray and the perfect specular direction. The first model is Lambertian,
\begin{equation}
f_r^{\rm L}={\rm const.},
\label{eq:mc_lambertian}
\end{equation}
which represents a broad diffuse scattering limit. The second is a Gaussian lobe around the specular direction,
\begin{equation}
f_r^{\rm G}(\Delta)\propto\exp\left[-\frac{\Delta^2}{2\sigma^2}\right],
\qquad \sigma=4^\circ,
\label{eq:mc_gaussian}
\end{equation}
and is used to represent a strongly quasispecular surface.

To describe a surface with a narrow core but substantially longer off specular tails we also use a power law of the form
\begin{equation}
f_r^{\rm PL}(\Delta)\propto\left[1+\left(\frac{\tan(\Delta/2)}{\tan(\theta_c/2)}\right)^2\right]^{-p},
\qquad \theta_c=3^\circ,\quad p=1.
\label{eq:mc_powerlaw}
\end{equation}
Finally, we consider a GGX microfacet model,
\begin{equation}
f_r^{\rm GGX}=\frac{D(\mathbf{h})F(\mathbf{w}_i,\mathbf{h})G(\mathbf{w}_i,\mathbf{w}_o)}{4\cos\theta_i\cos\theta_o},
\label{eq:mc_ggx}
\end{equation}
where $\mathbf{h}$ is the half vector between the incident and outgoing directions, $D$ is the microfacet normal distribution, $F$ the Fresnel factor, and $G$ the masking shadowing factor. We use a roughness parameter $\alpha_{\rm GGX}=0.25$ and normal incidence Fresnel reflectance $F_0=0.60$. This model provides an intermediate behavior in which the distribution retains a specular preference but possesses broad non Gaussian tails.

\begin{figure}[htbp]
    \centering
    \includegraphics[width=0.72\linewidth]{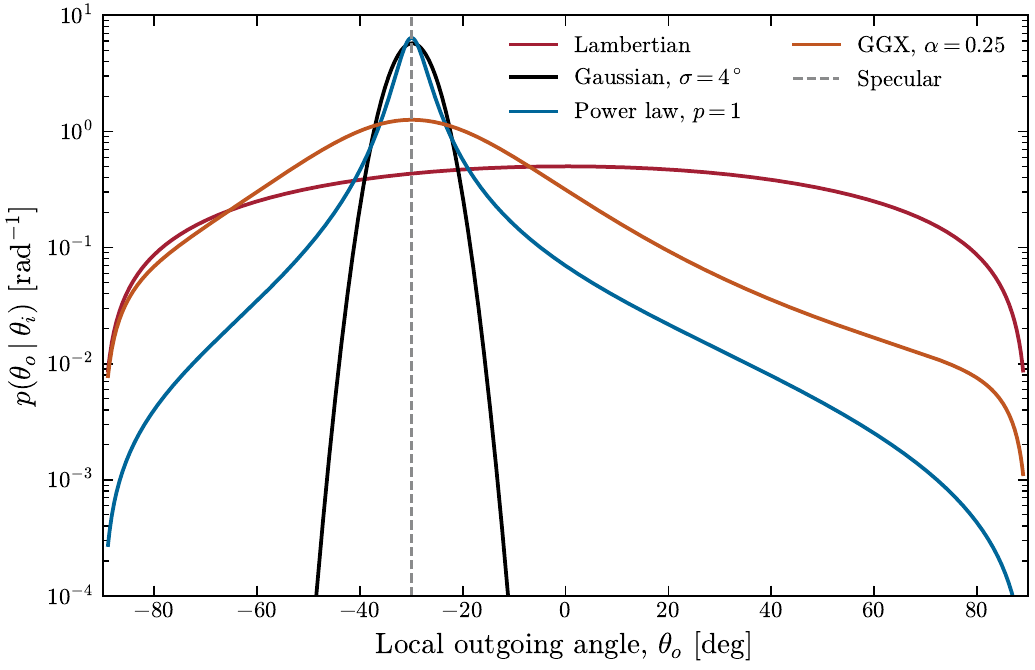}
    \caption{\justifying Representative local outgoing angle probability densities used for the beam dump Monte Carlo. The dashed line indicates the specular direction.}
    \label{fig:ghost_scattering_models}
\end{figure}

The different angular distributions are illustrated in Fig.~\ref{fig:ghost_scattering_models} for a representative angle of incidence. All models are normalized to the same hemispherical reflected fraction $R_{\rm bd}$, such that the figure compares only how the surviving reflected power is redistributed in angle. The Gaussian model concentrates the reflected power strongly around the specular direction, while the Lambertian distribution produces a broad diffuse return. The power-law model combines a narrow specular peak with comparatively long angular tails, whereas the GGX model provides an intermediate case with a broad specular lobe and substantial off-specular scattering. These differences explain why surfaces having the same total reflected fraction can nevertheless produce substantially different values of ${\rm BRDF}_{\rm bd}^{\rm eff}$ in the reverse ghost direction.

Representative BRDF measurements of candidate absorbing and low reflectivity surfaces show precisely this broad range of possible behavior~\cite{Ananyeva_BRDF,Ananyeva_BRDF_2,Kontos_BRDF} as some surfaces exhibit a very narrow specular core superimposed on a comparatively low off-specular background, whereas others retain substantially stronger wide angle scattering. Measurements of oxidized stainless steel, black glass, antireflection coated glass, multilayer antireflection coatings, and absorbing coatings can differ by orders of magnitude in the off-specular region and can also change significantly with angle of incidence~\cite{Ananyeva_BRDF,Ananyeva_BRDF_2,Kontos_BRDF}. The four distributions of Eqs.~\eqref{eq:mc_lambertian}--\eqref{eq:mc_ggx} are therefore intended as representative angular families rather than fits to any of these measurements.

This distinction is important because a small off-specular BRDF and a small total reflected fraction are separate desirable properties. The former controls the probability of returning into the reverse ghost direction at a single interaction, whereas the latter suppresses every subsequent interaction inside the dump. The Monte Carlo allows these two effects to be varied independently.

\subsection{Reflectivity and multiple bounce suppression}

For a fixed set of ray trajectories, changing $R_{\rm bd}$ simply changes the weight accumulated along each path. A ray undergoing $N$ reflections before returning contributes a factor proportional to $R_{\rm bd}^N$. The total effective BRDF can therefore be expressed schematically as
\begin{equation}
{\rm BRDF}_{\rm bd}^{\rm eff}(R_{\rm bd})=\sum_N C_N R_{\rm bd}^N,
\label{eq:mc_reflectivity_series}
\end{equation}
where $C_N$ contains the geometrical probability of the corresponding class of paths and its angular overlap with the reverse direction.

For the reference GGX configuration at $\theta_{\rm in}=10^\circ$, the Monte Carlo gives an effective scaling close to ${\rm BRDF}_{\rm bd}^{\rm eff}\propto R_{\rm bd}^{1.15}$ over the range considered, consistent with a mixture dominated by low-order reflections. By contrast, at the quasispecular retroreflection condition of the $50^\circ$ wedge, the returned light is dominated by the three-reflection path and follows ${\rm BRDF}_{\rm bd}^{\rm eff}\propto R_{\rm bd}^3$. This illustrates why reducing the surface reflectivity through an absorbing coating can be especially effective against geometrically resonant paths.

\begin{figure}[htbp]
    \centering
    \includegraphics[width=0.72\linewidth]{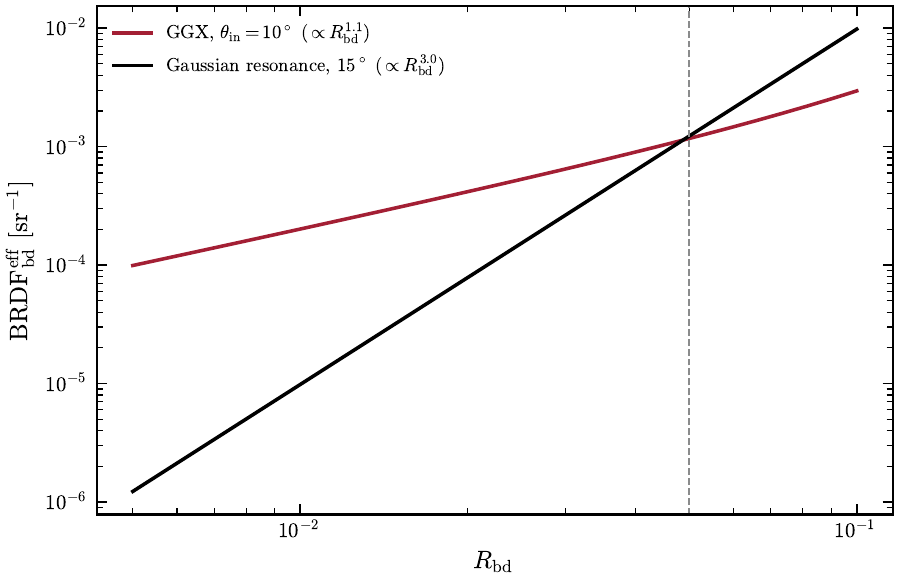}
    \caption{\justifying Dependence of the effective beam dump BRDF on the hemispherical reflected fraction $R_{\rm bd}$.}
    \label{fig:ghost_dump_reflectivity}
\end{figure}

\end{document}